\documentclass[aps,prl,twocolumn,amsmath,amssymb,superscriptaddress]{revtex4}
\pdfoutput=1
\usepackage{graphicx}
\usepackage{amssymb}
\usepackage{natbib}
\usepackage{calc}
\usepackage{SIunits}
\usepackage{textcomp}

\begin{document}

\title{Neutron spin resonance in a quasi-two-dimensional iron-based superconductor}

\author{Wenshan Hong}
\affiliation{Beijing National Laboratory for Condensed Matter
Physics, Institute of Physics, Chinese Academy of Sciences, Beijing
100190, China}
\affiliation{University of Chinese Academy of Sciences, Beijing 100049, China}
\author{Linxing Song}
\affiliation{Beijing National Laboratory for Condensed Matter
Physics, Institute of Physics, Chinese Academy of Sciences, Beijing
100190, China}
\affiliation{University of Chinese Academy of Sciences, Beijing 100049, China}
\author{Bo Liu}
\affiliation{Beijing National Laboratory for Condensed Matter
Physics, Institute of Physics, Chinese Academy of Sciences, Beijing
100190, China}
\affiliation{University of Chinese Academy of Sciences, Beijing 100049, China}
\author{Zezong Li}
\affiliation{Beijing National Laboratory for Condensed Matter
Physics, Institute of Physics, Chinese Academy of Sciences, Beijing
100190, China}
\affiliation{University of Chinese Academy of Sciences, Beijing 100049, China}
\author{Zhenyuan Zeng}
\affiliation{Beijing National Laboratory for Condensed Matter
Physics, Institute of Physics, Chinese Academy of Sciences, Beijing
100190, China}
\affiliation{University of Chinese Academy of Sciences, Beijing 100049, China}
\author{Yang Li}
\affiliation{Beijing National Laboratory for Condensed Matter
Physics, Institute of Physics, Chinese Academy of Sciences, Beijing
100190, China}
\affiliation{University of Chinese Academy of Sciences, Beijing 100049, China}
\author{Dingsong Wu}
\affiliation{Beijing National Laboratory for Condensed Matter
Physics, Institute of Physics, Chinese Academy of Sciences, Beijing
100190, China}
\affiliation{University of Chinese Academy of Sciences, Beijing 100049, China}
\author{Qiangtao Sui}
\affiliation{Beijing National Laboratory for Condensed Matter
Physics, Institute of Physics, Chinese Academy of Sciences, Beijing
100190, China}
\affiliation{University of Chinese Academy of Sciences, Beijing 100049, China}
\author{Tao Xie}
\affiliation{Beijing National Laboratory for Condensed Matter
Physics, Institute of Physics, Chinese Academy of Sciences, Beijing
100190, China}
\affiliation{University of Chinese Academy of Sciences, Beijing 100049, China}
\author{Sergey Danilkin}
\affiliation{Australian Centre for Neutron Scattering, Australian Nuclear Science and
Technology Organization, Lucas Heights NSW-2234, Australia}
\author{Haranath Ghosh}
\affiliation{Human Resources Development Section, Raja Ramanna Centre for Advanced Technology, Indore 452013, India}
\affiliation{Homi Bhabha National Institute, BARC training school complex, Anushakti Nagar, Mumbai 400094, India}
\author{Abyay Ghosh}
\affiliation{Human Resources Development Section, Raja Ramanna Centre for Advanced Technology, Indore 452013, India}
\affiliation{Homi Bhabha National Institute, BARC training school complex, Anushakti Nagar, Mumbai 400094, India}
\author{Jiangping Hu}
\affiliation{Beijing National Laboratory for Condensed Matter
Physics, Institute of Physics, Chinese Academy of Sciences, Beijing
100190, China}
\affiliation{University of Chinese Academy of Sciences, Beijing 100049, China}
\affiliation{Songshan Lake Materials Laboratory, Dongguan, Guangdong 523808, China }
\author{Lin Zhao}
\affiliation{Beijing National Laboratory for Condensed Matter
Physics, Institute of Physics, Chinese Academy of Sciences, Beijing
100190, China}
\affiliation{Songshan Lake Materials Laboratory, Dongguan, Guangdong 523808, China }
\author{Xingjiang Zhou}
\affiliation{Beijing National Laboratory for Condensed Matter
Physics, Institute of Physics, Chinese Academy of Sciences, Beijing
100190, China}
\affiliation{University of Chinese Academy of Sciences, Beijing 100049, China}
\affiliation{Songshan Lake Materials Laboratory, Dongguan, Guangdong 523808, China }
\author{Xianggang Qiu}
\affiliation{Beijing National Laboratory for Condensed Matter
Physics, Institute of Physics, Chinese Academy of Sciences, Beijing
100190, China}
\affiliation{University of Chinese Academy of Sciences, Beijing 100049, China}
\affiliation{Songshan Lake Materials Laboratory, Dongguan, Guangdong 523808, China }
\author{Shiliang Li}
\email{slli@iphy.ac.cn}
\affiliation{Beijing National Laboratory for Condensed Matter
Physics, Institute of Physics, Chinese Academy of Sciences, Beijing
100190, China}
\affiliation{University of Chinese Academy of Sciences, Beijing 100049, China}
\affiliation{Songshan Lake Materials Laboratory, Dongguan, Guangdong 523808, China }
\author{Huiqian Luo}
\email{hqluo@iphy.ac.cn}
\affiliation{Beijing National Laboratory for Condensed Matter
Physics, Institute of Physics, Chinese Academy of Sciences, Beijing
100190, China}
\affiliation{Songshan Lake Materials Laboratory, Dongguan, Guangdong 523808, China }

\date{\today}

\maketitle

{\bf
Magnetically mediated Cooper pairing is generally regarded as a key to establish the unified mechanism of unconventional superconductivity \cite{djscalapino2012,jttranquada2014,pdai2015}.  One crucial evidence is the neutron spin resonance arising in the superconducting state, which is commonly interpreted as a spin-exciton from collective particle-hole excitations confined below the superconducting pair-breaking gap ($2\Delta$) \cite{nksato2001,adhristianson2008,ysidis2007,meschrig2006,tdas2011}. Here, on the basis of inelastic neutron scattering measurements on a quasi-two-dimensional iron-based superconductor KCa$_2$Fe$_4$As$_4$F$_2$, we have discovered a two-dimensional spin resonant mode with downward dispersions, a behavior closely resembling the low branch of the hour-glass-type spin resonance in cuprates \cite{ysidis2007,meschrig2006}. The resonant intensity is predominant by two broad incommensurate peaks near $Q=$(0.5, 0.5) with a sharp energy peak at $E_R=16$ meV. The overall energy dispersion of the mode exceeds the measured maximum total gap $\Delta_{\rm tot}=|\Delta_k|+|\Delta_{k+Q}|$.  These experimental results deeply challenge the conventional understanding of the resonance modes as magnetic excitons regardless of underlining pairing symmetry schemes \cite{ysidis2007,meschrig2006,tdas2011,korshunov2008,parish2008,mgkim2013,sonari2010,hkontani2010,ltakeuchi2018}, and it also points out that when the iron-based superconductivity becomes very quasi-two-dimensional, the electronic behaviors are similar to those in cuprates.
}

Broadly speaking, there are currently only two families of unconventional superconductors that show high-$T_c$ superconductivity, i.e., cuprates and iron-based superconductors (FeSCs). Whether they share a common superconducting mechanism is still an open question at the frontier of modern condensed matter physics \cite{si2016}. Unlike the well-established intra-band $d-$wave pairing symmetry in cuprates (Fig. 1a), the iron-based superconductivity is complicated for its multi-band nature and can have sign-reversed $s^{\pm}$-wave pairings from inter-band repulsive interactions on the Fermi surfaces connected by a finite wavevector $Q$ \cite{prichard2011} (Fig. 1b).  Such pairing symmetry can be approached by two very different scenarios, either from the Fermi surface nesting between hole and electron pockets under weak coupling \cite{avchubukov2008a,tamaier2009,mazin2009}, or from short-range antiferromagnetic (AF) fluctuations under strong coupling similar to cuprates \cite{si2016,Seo2008,pjhirschfeld2011}. The latter picture is supported by the strong AF fluctuations throughout the phase diagram of FeSCs, and some endeavors have been applied in FeSCs to search for the Mott insulating state \cite{ysong2016a} and two-dimensional (2D) high-$T_c$ superconductivity in analogy to cuprates \cite{si2016}. Alternately, a conventional sign-preserved ($s^{++}$) pairing picture is also proposed in FeSCs, where the Cooper pairs are mediated by inter-band attractive interactions driven by orbital fluctuations \cite{sonari2010,hkontani2010} (Fig. 1c).

\begin{figure*}[t]
\includegraphics[width=0.85\textwidth]{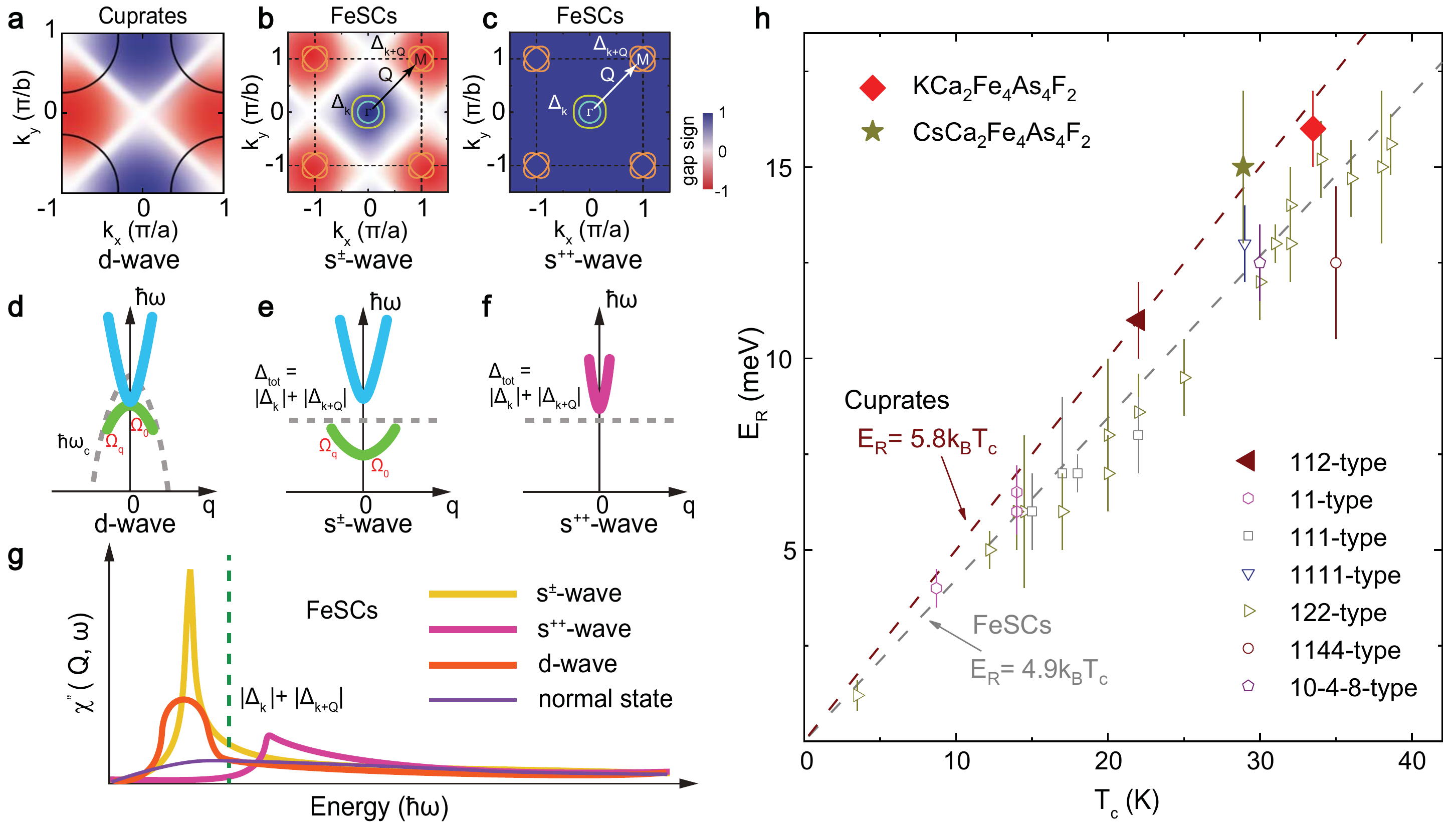}
\caption{
{\bf Neutron spin resonance and pairing symmetry in cuprates and iron-based high-$T_c$ superconductors.}\\
{\bf a}, Single-band $d-$wave pairing in hole-doped cuprates.
{\bf b}, Sign-reversed multi-band $s^{\pm}-$pairing in FeSCs.
{\bf c}, Sign-preserved multi-band $s^{++}-$pairing in FeSCs.
{\bf d, e, f}, Dispersion of spin excitations / resonance in the superconducting state corresponding to the pairing symmetries in {\bf a, b, c}.
{\bf g}, Comparison between the sharp resonant peak under $s^{\pm}-$pairing, broad resonant peak under $d-$pairing and broad spin-resonance-like hump under $s^{++}$-wave pairing in FeSCs as shown by the local susceptibility $\chi^{\prime\prime}(Q,\omega)$ below $T_c$.
{\bf h}, Summarized spin resonance energy $E_R$ versus $T_c$ in iron-based superconductors, where the grey dash line $E_R=4.9 k_BT_c$ scales most of them, and $E_R=5.8 k_BT_c$ is usually used in cuprates.
Here $\hbar\omega_c$ is the continuum threshold energy in cuprates, and $\Delta_{tot}=|\Delta_k|+|\Delta_{k+Q}|$ is defined as the total superconducting gaps summed on two Fermi surfaces connected by the wavevector $Q$.
}
\end{figure*}
In copper-oxide, heavy-fermion, iron-pnictide chalcogenide superconductors, a neutron spin resonance served as the hallmark of a sign-reversed pairing symmetry is intensively observed \cite{djscalapino2012,jttranquada2014,nksato2001,adhristianson2008,pdai2015,ysidis2007,meschrig2006,tdas2011}. This collective mode is born from the dynamic spin susceptibility $\chi^{\prime\prime}(Q,\omega)$ already present in the normal state, and its intensity changes with temperature like the superconducting order parameter \cite{inosov2010}. More importantly, the mode energy $E_R$ seems linearly related to either $T_c$ or the superconducting gap \cite{ysidis2007,gyu2009}, hinting a universal magnetic origin of the Cooper pairings in unconventional superconductors analogous to the phonon mediated conventional superconductivity \cite{meschrig2006,pmonthoux2007}. Although many theories have been proposed to explain the spin resonance in unconventional superconductors, the most successful candidate until now is the so-called spin-exciton picture \cite{meschrig2006,ysidis2007}. Namely, the spin resonance is a spin-1 exciton from the collective particle-hole excitations below a spin-flip continuum energy $\hbar\omega_c$, which should be slightly lower than twice of the superconducting gap (pair-breaking gap) $2\Delta$. Thus the dispersion of spin resonance is determined by the momentum dependence of the spin fluctuations and simultaneously limited by the continuum threshold $\hbar\omega_c$. In the single-band $d-$wave superconductors like cuprates, as the gap magnitude is strongly momentum dependent from the antinodal to nodal region, a downward dispersion of the resonance is observed beneath the dome of $\hbar\omega_c$ \cite{ysidis2007}. Together with the upward magnon-like dispersion at energies above $E_R$, they form an hour-glass-type spin excitations (Fig. 1d) \cite{jttranquada2014}. In the $s^{\pm}-$wave superconductors like FeSCs, $\hbar\omega_c$ is defined by the total superconducting gap summed on two Fermi surfaces linked by the wavevector $Q$: $\Delta_{\rm tot}=|\Delta_k|+|\Delta_{k+Q}|$, and it is usually momentum independent, the resonance should reveal an upward magnon-like dispersion instead (Fig. 1e) \cite{korshunov2008,parish2008,tdas2011,mgkim2013}. Alternately, a resonance-like hump is also proposed in FeSCs under the $s^{++}-$pairing picture due to the self-energy effect induced redistribution of spin fluctuations below $T_c$ \cite{sonari2010,hkontani2010,ltakeuchi2018}. When the enhancement of dynamical spin susceptibility from self energy exceeds the suppression due to the coherence factor effect in the superconducting state, the spin excitations may have a steep upward dispersion in momentum space (Fig. 1f) \cite{clzhang2013} and form a broad energy hump (Fig. 1g) above $\Delta_{\rm tot}$  \cite{hkontani2010,ltakeuchi2018}. For comparison, a broad spin resonant peak below $\Delta_{\rm tot}$ is expected when $d-$wave pairings emerge on the hole pockets in the zone center (Fig. 1g) \cite{tamaier2009}. Experimentally, only upward dispersions of the spin resonances are observed in FeSCs \cite{mgkim2013,clzhang2013,dhu2016,rzhang2018}, where its velocity is  normalized by Landau damping in comparison to the normal state spin fluctuations \cite{mgkim2013}, and $E_R$ is usually beneath $\Delta_{\rm tot}$ ($E_R/\Delta_{\rm tot}\approx 0.64$) \cite{gyu2009,txie2018b}, thus supporting the spin-exciton picture under $s^{\pm}-$pairing. Moreover, the linearly scaling ratio between $E_R$ and $k_BT_c$ among FeSCs is about 4.9, slightly lower than that in cuprates where $E_R/k_BT_c=5.8$ (Fig. 1h) \cite{meschrig2006,txie2018a}.
\begin{figure*}[t]
\includegraphics[width=0.85\textwidth]{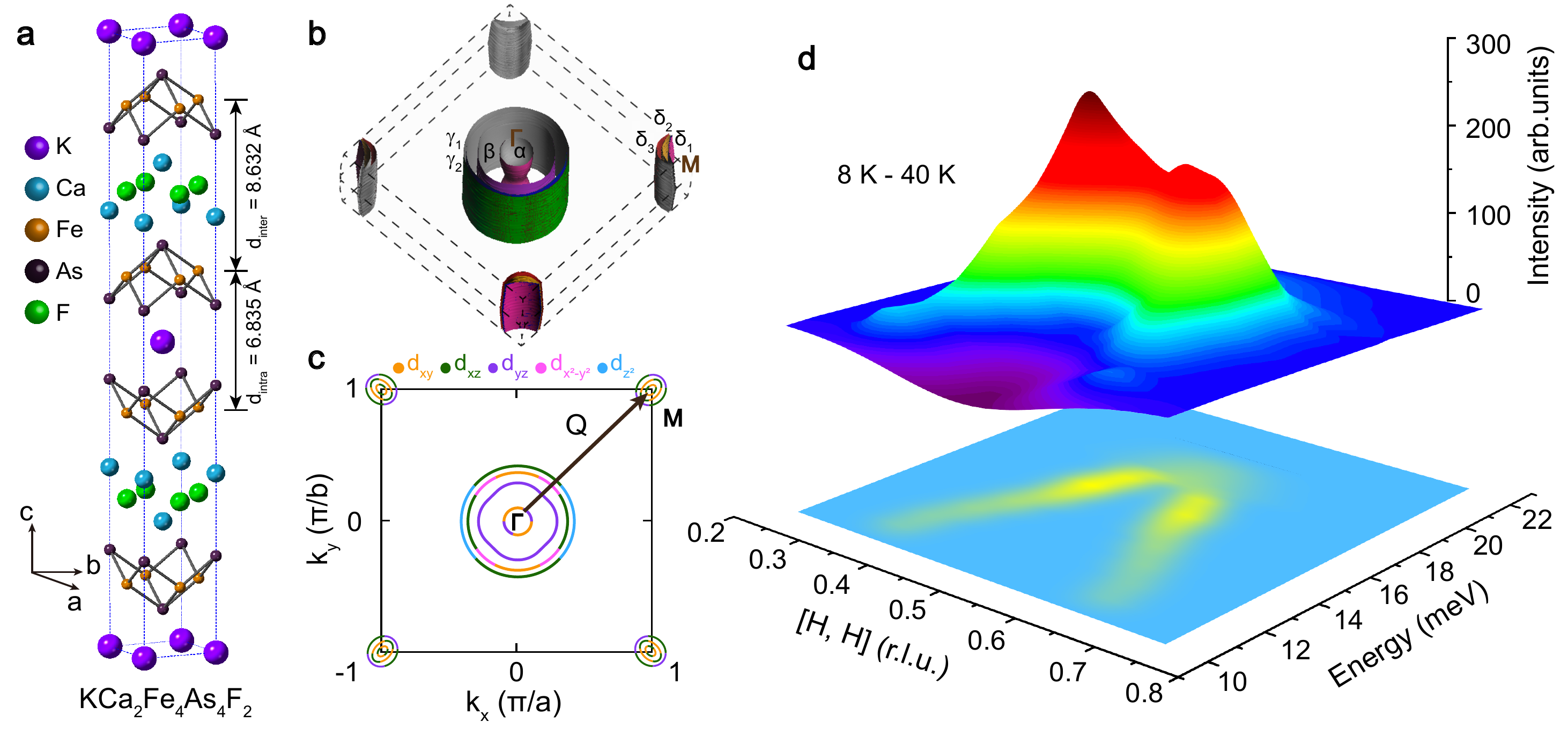}
\caption{
{\bf Crystal structure, Fermi surface and the spin resonance mode of KCa$_2$Fe$_4$As$_4$F$_2$}
{\bf a}, Crystal structure with interlaced stacks of CaFeAsF and KFe$_2$As$_2$, where the Fe$_2$As$_2$ bilayers are separated by insulating Ca$_2$F$_2$ blocks.
{\bf b} and {\bf c}, DFT calculation results on the Fermi surfaces\cite{supplementary}. There are 4 visible hole pockets with distinct sizes around $\Gamma$ point ($\alpha, \beta, \gamma_1, \gamma_2$) and 3 small electron pockets ($\delta_{1,2,3}$) around $M$ point. The Fermi pockets corresponding to are shown in the 2D Brillouin zone of 2-Fe unit cell, and each of them consists of different orbitals of Fe$^{2+}$ such as $d_{xy}$, $d_{xz}$, $d_{yz}$, $d_{x^2-y^2}$ and $d_{z^2}$ .
{\bf d}, Sketch picture of the spin resonance mode and its downward dispersion. The intensity are obtained from the neutron counts by subtracting the 40 K data at normal state from 8 K data in the superconducting state.
}
\end{figure*}

The newly discovered 12442-type FeSC KCa$_2$Fe$_4$As$_4$F$_2$ with $T_c=33.5$ K exhibits many properties strongly resembling those in cuprates. Its crystal structure can be viewed as an inter-growth of 1111-type CaFeAsF and 122-type KFe$_2$As$_2$, where the asymmetric bilayers of Fe$_2$As$_2$ are separated by the insulating Ca$_2$F$_2$ layers (Fig. 2a) \cite{zcwang2016a}, much like the double CuO$_2$ planes structure in copper oxides like Bi$_2$Sr$_2$CaCu$_2$O$_{8+\delta}$ (Bi2212), YBa$_2$Cu$_3$O$_{6+\delta}$ (YBCO) and La$_{2-x}$Sr$_x$CaCu$_2$O$_6$ (LSCCO), etc \cite{braveau2013}. Because the distance of intra-bilayer $d_{intra}=6.835$ \AA\ is similar to the interlayer spacing in KFe$_2$As$_2$ ($d=6.94$ \AA), and the inter-bilayer distance is even larger with $d_{inter}=8.632$ \AA\ , both of them are not close enough to establish a 3D long-range AF order \cite{zcwang2016b}. Thus this compound is naturally paramagnetic in a body-centred tetragonal lattice, and such structure yields a hole-type self-doping at a level of 0.25 holes/Fe without any impurity effects from chemical dopants \cite{gtwang2016b}. Transport measurements on 12442-type compounds suggest that both normal state and superconducting state are highly anisotropic \cite{abyu2019,zcwang2019,twang2019,twang2020}, much like the quasi-2D characteristic in cuprates. Our recent angle-resolved-photoemission-spectroscopy (ARPES) measurements also reveal a clear band splitting effect similar to the case in Bi2212, but here it is induced by the interlayer interorbital interactions within the Fe$_2$As$_2$ bilayer unit \cite{meschrig2006,dswu2020}.  While the Density-Functional-Theory (DFT) calculations based on $ab-initio$ electronic structures suggest 10 degenerated bands forming four visible hole pockets with distinct sizes around $\Gamma$ point and three small electron pockets around $M$ point (Fig. 2b) \cite{supplementary}, three splitting hole pockets and a very tiny electron pocket ($r_\delta\approx0.04 \pi/a$) are also observed by ARPES measurements \cite{dswu2020}. All these facts suggest the Fermi surface nesting along $(\pi, \pi)$ direction is unlikely to happen in this compound. However, the $s^{\pm}$-wave pairings under strong coupling approach can still occur between the hole and electron pockets connected by a longitudinal wavevector $Q$ along $(\pi, \pi)$ direction (Fig. 2c), forming a spin resonant mode around wavevector $Q$.

\begin{figure*}[t]
\includegraphics[width=0.95\textwidth]{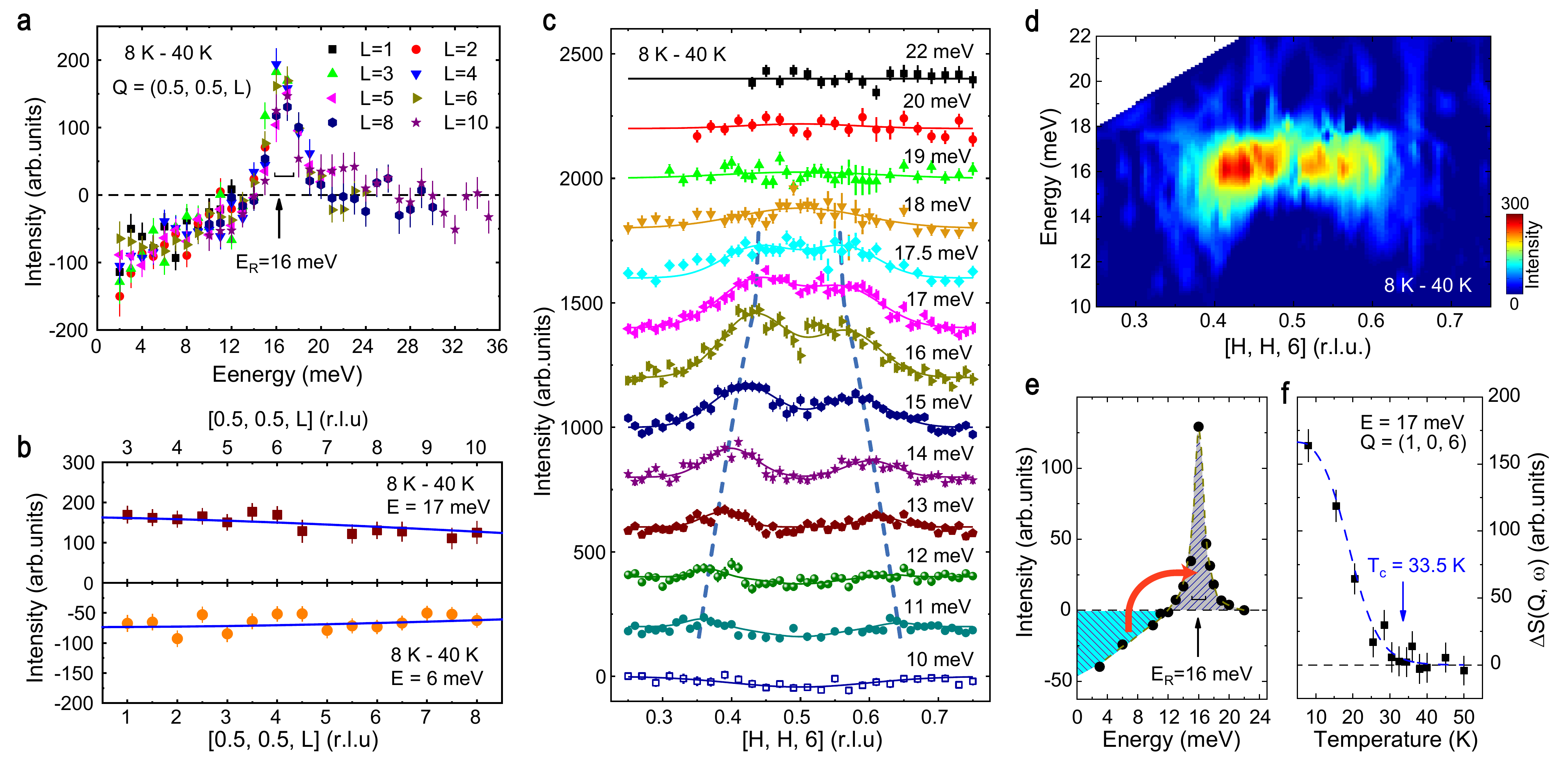}
\caption{{\bf Neutron spin resonant mode in KCa$_2$Fe$_4$As$_4$F$_2$.}
{\bf a}, Spin resonant peak as shown by the difference between $T=8$ K and $T=40$ K of energy scans at $Q=(0.5, 0.5, L)$ with $L=1 \sim 10$.
{\bf b}, $L-$dependence of the difference between $T=8$ K and $T=40$ K of spin excitations at $E=17$ meV and 6 meV, respectively. The solid lines represent the square of magnetic form factor $\mid F(Q) \mid^2$ after normalizing to the intensity.
{\bf c}, Constant-energy scans of the intensity difference between $T=8$ K and $T=40$ K along the $[H, H, 6]$ direction from $E=10$ meV to 22 meV. All data are shifted by a fixed step for clarity. The solid lines are two-gaussian-peak fitting curves, where the data from 18 meV to 20 meV is only fitted by single gaussian peak.
{\bf d}, Color mapping of the resonant intensity obtained from c.
{\bf e}, Integrated intensity differences between $T=8$ K and $T=40$ K from 3 meV to 22 meV. The shadow regions suggest similar areas for the negative and positive intensities.
{\bf f}, Temperature dependence of the spin resonance intensity at $E=17$ meV and $Q=(0.5, 0.5, 6)$ after subtracting a constant background from 35 to 50 K. The arrow marks the $T_c=33.5$ K.
Here the horizontal bar shows the energy resolution of Taipan spectrometer, and all dashed lines are guides to eyes.
 }
 \end{figure*}

\begin{figure}[t]
\includegraphics[width=0.4\textwidth]{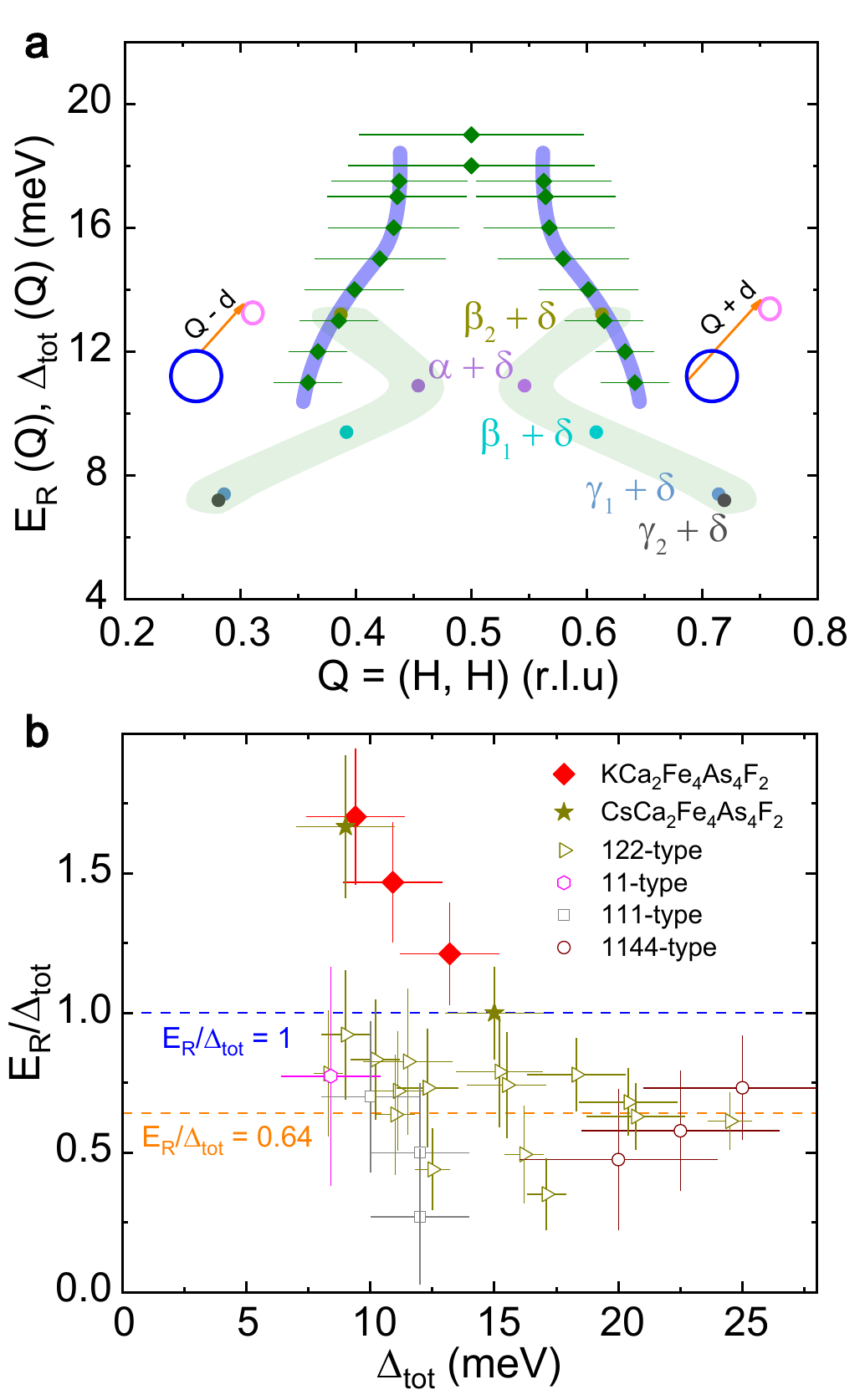}
\caption{{\bf Comparison between the resonant energy $E_R$ and the total superconducting gap $\Delta_{tot}$. }
{\bf a}, Dispersion of the spin resonance above $\Delta_{tot}(Q)$ in KCa$_2$Fe$_4$As$_4$F$_2$. The gaps are summed on two Fermi pockets linked by longitudinally incommensurate wave vectors $Q+d$ and $Q-d$. The horizontal bars are the peak width obtained from the gaussian fittings in Fig. 3c. The shadow areas mark the size of electron pocket at $M$ point \cite{supplementary}.
{\bf b}, The ratio of $E_R/\Delta_{tot}$ in 12442-type FeSCs and other compounds. Here we use three values of $\Delta_{tot}$ for KCa$_2$Fe$_4$As$_4$F$_2$ ($\Delta_{tot}=$ 13.2 meV ($\beta_2+\delta$), 10.9 meV ($\alpha+\delta$), 9.4 meV ($\beta_1+\delta$) from ARPES results), and two values for CsCa$_2$Fe$_4$As$_4$F$_2$ ($\Delta_L+\Delta_L$= 15 meV and $\Delta_L+\Delta_S$= 9 meV from $\mu$SR results).
 }
\end{figure}

Here using inelastic neutron scattering to study the low-energy spin excitations of KCa$_2$Fe$_4$As$_4$F$_2$ single crystals, we have discovered a spin resonant mode at 16 meV with two incommensurate peaks and downward dispersions along $[H, H]$ direction (Fig. 2d). The main results are shown in Fig. 3.  We have performed energy scans at $Q=(0.5, 0.5, L)$ from 2 meV to 35 meV for L from 1 to 10 in the $[H, H, L]$ scattering plane. After subtracting the intensity of spin excitations in the normal state ($T=40$ K), we can identify a spin resonant mode in the superconducting state at $T=8$ K for a clear spectral-weight gain above 13 meV and a depletion below this energy. The resonant peak has a maximum intensity at $E_R=$ 16 meV and a nearly resolution-limited width, and all data are overlapped for different $L$s (Fig. 3a). Further $L-$dependence were measured at fixed energies $E=17$ meV and 6 meV, both of which show no $L-$ modulation but simply follow the square of Fe$^{2+}$ magnetic form factor (Fig. 3b). Thus the resonance is completely 2D, consistent with the transport results for the superconductivity confined within the Fe$_2$As$_2$ bilayers \cite{zcwang2019,twang2019,abyu2019,twang2020}. To map out the dispersion of the resonance, we have carried out systematic constant-energy scans along $[H, H, 6]$ direction from 3 meV to 22 meV. The contamination of the phonon scattering can be removed by the intensity difference between 8 K and 40 K, which clearly reveals two incommensurate resonant peaks along $[H, H]$ direction in broad widths (Fig. 3c). There is a weak negative intensity difference below 12 meV, but two asymmetric peaks emerge above 11 meV and their intensities quickly increase before finally merge together at 18 meV then disappear above 20 meV. We demonstrate the spectral-weight gain for the resonance mode in 2D color mapping in Fig. 3d, where most contributions are from spin excitations around 16 meV, namely, the resonant energy $E_R$. After applying a two-gaussian-peak fitting, we obtain an explicit downward dispersion of the resonance with increasing peak width upon increasing energy (Fig. 3c). At 18 meV and 19  meV, we have to use a single gaussian peak function to fit for the broad and weak intensity. The energy dependence of the integrated intensity from fitting functions is present in Fig. 3e. The resonant peak is even sharper compared to that in the energy scans and clearly resolution limited. The spectral-weight is conserved for nearly identical areas of the negative and positive parts.  Fig. 3f shows the order-parameter-like intensity gain for the resonance at $E=17$ meV and $Q=(0.5, 0.5, 6)$,  which decreases upon warming up and ceases at $T_c=33.5$ K.

We further summarize the dispersion of spin resonance and compare with the momentum dependence of $\Delta_{\rm tot}(Q)$ from ARPES results \cite{dswu2020} in Fig. 4a. Since both the pocket sizes and superconducting gaps are quite divergent ($\Delta_{\alpha}=5.6$ meV, $\Delta_{\beta1}=4.1$ meV, $\Delta_{\beta2}=7.9$ meV, $\Delta_{\gamma1}=2.1$ meV, $\Delta_{\gamma2}=1.9$ meV and $\Delta_{\delta}=5.3$ meV), and the hole pockets are strongly mismatched with the only electron pocket \cite{supplementary} ($r_h >> r_e$), $\Delta_{\rm tot}$ actually has different magnitudes at several incommensurate wavevetors $Q \pm d$ along $[H, H]$ direction, where the incommensurability $d$ is determined by the mismatched radiuses $\sqrt{r_h^2-r_e^2}$. Although an hour-glass-like shape of $\Delta_{\rm tot}(Q)$ qualitatively scales with the downward dispersion of the spin resonance, the overall resonant intensities are clearly above $\Delta_{\rm tot}$ for most of $Q$ positions, where we couldn't identify the incommensurability of those resonant peaks above 18 meV for their weak intensities and broadening widths. Such results are apparently inconsistent with the spin-exciton picture under $s^{\pm}-$pairing \cite{tamaier2009,tdas2011,mgkim2013}. The resolution limited peak width of energy dependence together with downward dispersions cannot be explained by the $s^{++}-$pairing picture \cite{sonari2010,hkontani2010,ltakeuchi2018}, either. Only when the system is proximate to the magnetic quantum critical point (QCP), $s^{++}-$pairing may give a sharp peak of $\chi^{\prime\prime}(Q,\omega)$ above $\Delta_{\rm tot}$  \cite{ltakeuchi2018}. However, KCa$_2$Fe$_4$As$_4$F$_2$ has a hole concentration similar to the overdoped Ba$_{1-x}$K$_x$Fe$_2$As$_2$ and certainly far away from a magnetic instability or QCP \cite{zcwang2016a,zcwang2016b,gtwang2016b}, as electron dopings from Co or Ni substitutions cannot induce any magnetic orders but only suppress the $T_c$ \cite{jishida2017}.  We give a direct comparison between the resonant energy $E_R$ and $\Delta_{\rm tot}$ together with other FeSCs in Fig. 4b, here we use $\Delta_{\rm tot}=$ 13.2 meV ($\beta_2+\delta$), 10.9 meV ($\alpha+\delta$), 9.4 meV ($\beta_1+\delta$), respectively. For all cases, we have $E_R/\Delta_{\rm tot}> 1$. Neutron scattering on the powder sample of another 12442-type FeSC CsCa$_2$Fe$_4$As$_4$F$_2$ ($T_c=28.9$ K) suggests a spin resonance at $E_R=$ 15 meV \cite{dtadroja2020}, and $\mu$SR experiments give two superconducting gaps $\Delta_L$= 7.5 meV and $\Delta_S$= 1.5 meV, resulting in $E_R/\Delta_{\rm tot}\geq 1$, too.  Apparently, the large ratio of $E_R/\Delta_{\rm tot}$ makes the 12442-type FeSCs different from other systems, where the scaling of $E_R/\Delta_{\rm tot}= 0.64$ holds for most compounds.

Regardless of the abnormal ratio of $E_R/\Delta_{\rm tot}$, the observed spin resonance in KCa$_2$Fe$_4$As$_4$F$_2$ at $Q=(0.5, 0.5)$ (or $(\pi, \pi)$) certainly cannot be explained by the Fermi surface nesting picture but basically agrees with the strong coupling picture \cite{dswu2020}, and it shares many commonalities with cuprates  \cite{djscalapino2012,ysidis2007,meschrig2006}. Firstly, the resonant energy $E_R=16$ meV gives a ratio $E_R/k_BT_c$ of 5.5, larger than those other FeSCs (Fig. 1h) \cite{txie2018a}. Together with CsCa$_2$Fe$_4$As$_4$F$_2$ and the 112-type FeSC, such ratio in these materials with 2D spin resonance seems to follow the relation $E_R/k_BT_c=5.8$ in cuprates \cite{gyu2009,txie2018a,dtadroja2020}. Secondly, the 2D nature of the resonant intensity directly response to the anisotropic superconductivity similar to that in cuprates  \cite{meschrig2006,abyu2019,zcwang2019,twang2019,twang2020}. Due to weak intra-bilayer magnetic interactions, a splitting for the odd and even $L-$modulated resonant modes as found in CaKFe$_4$As$_4$, underdoped YBCO and Bi2212 does not appear in this compound \cite{ysidis2007,txie2018b}. Finally, the downward dispersion of the spin resonance mode intimately resembles the lower branch of the hour-glass-type of spin excitations in hole-doped cuprates \cite{jttranquada2014,meschrig2006,ysidis2007}. It should be noticed that a small $d-$wave gap ($<2$ meV) may exist in KCa$_2$Fe$_4$As$_4$F$_2$ as shown by $\mu$SR experiments \cite{msmidman}, but it can't cause a downward dispersion of the spin resonance at high energy.  For reference, in the heavily hole overdoped KFe$_2$As$_2$ with possible line nodes in the gaps, the spin excitations are incommensurate both at normal and superconducting state \cite{chlee2011,sdsheng2020}. Furthermore, in a heavy-fermion compound Ce$_{1-x}$Yb$_x$CoIn$_5$ with $d-$wave pairings on multiple fermi surfaces, an upward-dispersing resonance mode has been revealed, which is argued to be induced by the strong couplings with the 3D spin waves through the hybridization between $f$ electrons and conduction electrons \cite{ysong2016}. Therefore, along with our results in KCa$_2$Fe$_4$As$_4$F$_2$, these counter examples suggest the spin-exciton scenario of the spin resonance may not be appropriate, when the local moments and itinerant electrons are strongly coupled in the multi-band unconventional superconductors. Further theoretical and experimental investigations on the origin of the spin resonance are highly desired in these related compounds concerning the dimensionality of electronic behaviors, and it will certainly inspire the quest for an universal mechanism of the magnetically driven picture of unconventional superconductivity.

\begin{flushleft}
{\bf Methods}
The single crystals of KCa$_2$Fe$_4$As$_4$F$_2$ were grown by self-flux method \cite{twang2019,twang2020}. The crystals are in good quality for highly $c-$orientated reflection peaks in X-ray diffraction (XRD) measurements, and sharp superconducting transitions at $T_c=33.5$ K in resistivity and magnetization measurements. We examined the crystals piece by piece to exclude the KFe$_2$As$_2$ and CaFeAsF impurity phases by XRD, and finally obtained about 2.1 grams ($\sim$ 950 pieces) of high pure single crystals, and co-aligned them by hydrogen-free glue on aluminum plates using an optical microscopes and an X-ray Laue camera \cite{supplementary}.

Neutron scattering experiments were carried out using thermal triple-axis spectrometer Taipan at Australian Centre for Neutron Scattering, ANSTO, Australia, where the final neutron energy was fixed as $E_f=$ 14.8 meV, with a pyrolytic graphite filter, a double focusing  monochromator and a vertical focusing analyzer. The scattering plane $[H, H, 0] \times [0, 0, L]$ was defined by ${\bf Q}=(H,K,L) = (q_xa/2\pi, q_yb/2\pi, q_zc/2\pi)$ in reciprocal lattice unit (r.l.u.) using the tetragonal lattice: $a=b= 5.45$ \AA, $c=30.02$ \AA. The total mosaic of our sample mount was about 4.5$^{\circ}$ both in $[H, H, 0]$ and $[0, 0, L]$ directions. All raw data for inelastic neutron scattering measurements are given in the supplementary materials \cite{supplementary}.

Plane wave pseudopotential based DFT calculations were performed using Quantum Espresso numerical code \cite{pgiannozzi2009}. The plane wave cut off energy is set to 35 Ry after performing convergence test. Self-consistent field calculations are performed on 10$\times$10$\times$10 grid in momentum space. The Fermi surfaces are estimated using Cambridge Serial Total Energy Package using Generalized gradient approximation and Perdew-Burke-Enzerhof  exchange correlation \cite{sjclark2005}. In the latter case, the cut off energy and $k$ mesh are 500 eV and 26$\times$26$\times$37, respectively. We have used experimental lattice parameters \cite{zcwang2016a,zcwang2016b} as input to perform the single point energy calculations.
\end{flushleft}

\begin{flushleft}
{\bf Acknowledgements}
 This work is supported by the National Key Research and Development Program of China (2018YFA0704200, 2017YFA0303100, 2017YFA0302900 and 2016YFA0300500), the National Natural Science Foundation of China (11822411, 11961160699, 11874401, 11674406 and 11674372), the Strategic Priority Research Program (B) of the Chinese Academy of Sciences (CAS) (XDB07020300, XDB25000000), and Beijing Natural Science Foundation (JQ19002). H. L. and L. Z. are grateful for the support from the Youth Innovation Promotion Association of CAS (2016004,2017013). A. G. acknowledges HBNI RRCAT for financial support. This work is based on experiments performed at the Australian Centre for Neutron Scattering, ANSTO under a user program (Proposal No. P7795).
\end{flushleft}

\begin{flushleft}
{\bf Author Contributions}
 W. H., L. S., B. L., Z. L., Z. Z. and Y. L. grew the crystals of KCa$_2$Fe$_4$As$_4$F$_2$. W. H., L. S., B. L., D. W., Q. S., T. X. and X. Q. performed the sample characterizations and alignments. H. L., S. D. and W. H. performed the neutron scattering experiments. H. G. and A. G. did the DFT calculations. J. H. gave theoretical suggestions. D. W., L. Z. and X. Z. provided the ARPES results. W. H., S. L. and H. L. analysis the data and wrote the manuscript. The project was supervised by S. L. and H. L. All authors discussed the results, interpretation and conclusion.
\end{flushleft}

\begin{flushleft}
{\bf Competing interests} The authors declare no competing interests.
\end{flushleft}

\begin{flushleft}
{\bf Additional information} Correspondence and requests for materials should be addressed to S. L. (email: slli@iphy.ac.cn) or H. L. (email: hqluo@iphy.ac.cn).
\end{flushleft}

\clearpage
\appendix
\section{Supplementary Materials}
\begin{center}
{\bf A. SAMPLE GROWTH AND CHARACTERIZATION}
\end{center}

High quality single crystals of KCa$_2$Fe$_4$As$_4$F$_2$ were grown using self-flux method according to previous reports \cite{zcwang2016as,zcwang2017s,twang2019s,zcwang2019s}. KCa$_2$Fe$_4$As$_4$F$_2$ belongs to the 12442-type iron-based superconductors \cite{hjiang2013s,zcwang2017bs,zcwang2017cs,sqwu2017s}, its crystal structure is an intergrowth of the 1111-type CaFeAsF and 122-type KFe$_2$As$_2$ iron-based superconductors \cite{zcwang2016as,zcwang2017s,hjiang2013s}, very similar to the 1144-type superconductors CaKFe$_4$As$_4$ \cite{aiyo2016s,wmeier2017s} and EuRbFe$_4$As$_4$ \cite{jkbao2018s}. As the crystallization temperature for 122-type compounds are closed to 1144-type and 12442-type compounds, a precise control of the furnance temperature when cooling down the melted mixtures are critical in the crystal growth process. Even though, the formation of the impurity phases of CaFeAsF and KFe$_2$As$_2$ is very likely occurring during the sample growth. To obtain high pure samples, we must do detail sample characterizations and check their qualities. Here, we present the transport measurements, X-ray diffraction(XRD) and elastic neutron scattering results on our samples.

Figure S1a and S1b show the photos of our samples used in neutron experiments. The typical sizes of our crystals are about 3$\sim$5 mm with cleave surface on $ab-$plane. The thickness along $c-$axis is very thin ($<0.2$ mm). Using an X-ray Laue camera and an optical microscope, we can easily determine the crystal orientation by examining the 4-fold reflection patterns and some typical cleave lines. We have co-aligned about 950 pieces of single domain cystals in a total mass about 2.1 grams on thin aluminium plates by \emph{CYTOP} hydrogen-free glue in $[\emph{H},\emph{H},0] \times [0,0,\emph{L}]$ scattering plane (Fig. S1(b)).

The x-ray diffraction(XRD) was performed on an x-ray diffractometer \emph{SmartLab} 9 kW high resolution diffraction system with Cu K$_{\alpha}$ radiation($\lambda$ = 1.5406 \angstrom) at room temperature ranged from $5\degree$ to $80\degree$ in reflection mode. Typical XRD patterns are shown in Fig. S1c. Only sharp peaks along (00 $L$) ($L=$ even) orientation can be observed, suggesting high $c-$axis orientation and high quality of our samples. Figure S2 shows the temperature dependence of resistivity and magnetic susceptibility on a typical crystal. No anomaly features were found within the measured temperature range, suggesting the absence of antiferromagnetism transition. The superconducting transition temperature is about $T_c=33.5$ K, as indicated by sharp transitions on zero-resistivity and fully diamagnetic signal, which is closed to previous reports \cite{zcwang2016as,zcwang2017s,twang2019s,zcwang2019s}. If there is impurity phases of CaFeAsF, the resistivity should show a kink around 121 K \cite{fhan2008s,mtegel2008s,xyzhu2009as,xyzhu2009bs,smatsuishi2009s,yxiao2009s}. And if there is impurity phases of KFe$_2$As$_2$, further diamagnetic transition at $T_{c2}\approx3$ K \cite{hchen2009s,jkdong2010s} will be detected. Moreover, as the $c-$axis parameter of KCa$_2$Fe$_4$As$_4$F$_2$ (about 30 \AA) \cite{zcwang2016as,zcwang2017s} is much longer than both of them (about 8.5 \AA\ for CaFeAsF \cite{smatsuishi2009s,yxiao2009s} and 13.88 \AA\ for KFe$_2$As$_2$ \cite{hchen2009s}, respectively), the single crystal XRD would be a convenient tool to separate them as well. All crystals used for neutron scattering experiments are carefully selected piece by piece to make sure the impurity phases of CaFeAsF or KFe$_2$As$_2$ as less as possible.

\begin{center}
{\bf B. RAW DATA OF NEUTRON SCATTERING EXPERIMENTS}
\end{center}

Neutron scattering experiments were carried out using thermal triple-axis spectrometer Taipan at Australian Centre for Neutron Scattering, ANSTO, Australia, where the final neutron energy was fixed as $E_f=$ 14.8 meV, with a pyrolytic graphite filter, a double focusing  monochromator and a vertical focusing analyzer. The scattering plane $[H, H, 0] \times [0, 0, L]$ was defined by ${\bf Q}=(H,K,L) = (q_xa/2\pi, q_yb/2\pi, q_zc/2\pi)$ in reciprocal lattice unit (r.l.u.) using the tetragonal lattice: $a=b= 5.45$ \AA, $c=30.02$ \AA. The resolution of Taipan in the energy range we measured is about 1$\sim$2 meV, and specifically is about 1.8 meV for $E=16$ meV.

Before the inelastic measurements, we firstly performed elastic neutron scattering measurements to check the sample mosaic and quality. Fig. S3a and S3b show that the sample mosaic is about $4.5\degree$ for (0, 0, 4) Bragg peak and $4.4\degree$ for (2, 2, 0) Bragg peak, respectively. Assuming there are some impurity phases of CaFeAsF and KFe$_2$As$_2$ with same $c-$orientation as all samples are very thin, we can check them by a broad rocking curve measurements ($S1$ scan) when fixing the scattering angle $2\theta$ ($S2$ angle in spectrometer definition) at their reflections (0, 0, 1) or (0, 0, 2) with different lattice parameters. Such peaks are away from the first peak of KCa$_2$Fe$_4$As$_4$F$_2$ ($Q=$(0, 0, 2) with $S1\approx0\degree$ and $S2=-8.7\degree$ ), if they exist. In our measurements, both peaks were not found by scanning $S1$ from $-20\degree$ to $+20\degree$. Therefore, the bulk sample mount is dominated by pure KCa$_2$Fe$_4$As$_4$F$_2$ phase. We also check the $L-$modulation of the Bragg peaks at $Q=$(0, 0, $L$), (0.5, 0.5, $L$) and (1, 1, $L$), the $L=$ even nuclear scattering is very clear but broad in peak width, and at $L=7$, there possibly is a supurious peak from aluminium sample holder (Fig. S3 e, f, g). For (0.5, 0.5, $L$) scan, it is flat and just incoherent backgrounds, also suggesting no stripe-type \cite{pdai2012s,pdjohnsons} or spin-vortex-type \cite{wreier2018s,akreyssig2018s} antiferromagnetic orders in our sample, which is consistent with previous reports \cite{zcwang2016as,zcwang2017s}.

Figure S4 summarizes the raw data of energy scans at $Q = (0.5, 0.5, L)$ ($L$ = 1, 2, 3, 4, 5, 6, 8, 10). The signal grows quickly at low energy when close to the elastic tail. A strong peak appears around 16 meV for almost all scans that can be accounted for the phonon scattering of aluminum holders involved with the magnetic scattering from the samples. As the phonon signal wouldn't change within the temperature range 8 - 40 K, the neutron spin resonance can be identified by the intensity gain after subtracting the raw data above $T_c$ from the data below $T_c$ (Fig. 3a).

Figure S5 shows constant-energy scans along $L$ direction at fixed $E$ = 6 meV and 17 meV. For $E$ = 6 meV, the intensity both at 8 K and 40 K decreases as $L$ increases, which is roughly in compliance with the Fe$^{2+}$ form factor. While at $E$ = 17 meV, the intensity raised quickly at large $L$ due to the enhancement of phonon excitations at large $Q$. The $L-$modulation of the resonant mode thus can be also obtained by the different intensity between 8 K and 40 K (Fig. 3b).

In order to figure out the momentum dependence of the spin resonant peak, we have carried out measurements over a wide range of $E$ ($E$ = 6, 10, 11, 12, 13, 14, 15, 16, 17, 17.5, 18, 19, 20, 22 meV) along $[H, H, 6]$ direction, as shown in Fig. S6. The data for $E=3$ meV is similar to 6 meV case thus it is not shown here. The phonon supurious persist in almost all these $Q-$scans, and increases upon energy when disperse to high $Q$ positions. At the energies where $E$ is well above the resonant energy $E_R=$ 16 meV, such as $E$ = 19, 20, 22 meV, this strong peak at 8 K and 40 K has almost the same counts for its nature of phonon excitations. While around the resonant energy, clearly differences can be seen when cooling down below superconducting temperature. The weak peaks marked at the data both above and below $T_c$ represent the magnetic scattering signal in normal state and superconducting state, suggesting incommensurate features in both states.  After subtracting the 40 K data from 8 K data, we thus plot the spin resonance along $[H, H]$ direction for each energy (Fig. S6 b1-b14). We have used double-Gaussian-peak function to perform the fittings for $E$ = 12 $\sim$ 17.5 meV, where
\begin{center}
\begin{equation}
{\Delta S(Q, \omega) = \dfrac{A_1}{w\sqrt{{\pi}/2}}e^{{-2}{\frac{(H-0.5-d)}{w^{2}}^{2}}} + \dfrac{A_2}{w\sqrt{{\pi}/2}}e^{{-2}{\frac{(H-0.5+d)}{w^{2}}^{2}}}}.
\end{equation}
\end{center}
Here we simply suppose the two peaks have same width, but different intensities due to the normalization effect from Fe$^{2+}$ form factor and instrument setup. the parameter $d$ is the incommensurability. For those energies where the curves display only one broad peak, we use single peak Gaussian function to fit the data. As $E$ decreases from $E$ = 20 meV to 11 meV, the commensurate resonant peak splits into two incommensurate peaks with larger $d$ at the lower energies. The Full-With-at-Half-Maximum(FWHM) increases when $E$ moves upon energy. When $E< 12$ meV, the net intensity becomes negative around $Q=$(0.5, 0.5) first and spread to the entire zone at $E=6$ meV.

We also check the $L-$dependence of incommensurability by measuring $Q-$scans at 12 meV both for $L=3$ and $L=6$, as shown in Fig. S7. The spin resonance shows same incommensurate peaks at similar positions, but the intensity is different for the normalization effect from magnetic form factor.

Figure S8 shows the contour plot of $[H, H]$ scan both below and above $T_c$. Strong phonon excitations can be clearly seen at both temperatures. While in the region $E=$ 14 - 19 meV and $H$ = 0.3 - 0.7, intensity difference can be found for the spin resonant signal below $T_c$ (Fig. 3d).

\begin{center}
{\bf C. SUPERCONDUCTING GAPS AND FERMI SURFACES}
\end{center}

The iron-based superconductors are multiband systems. Multiple hole-like Fermi pockets around $\Gamma$ point are separated from the electron-like Fermi pockets around $M$ point, and superconducting gaps usually have different magnitudes (probably different signs) on each pocket \cite{prichard2011s}. For a fixed $k_z$ plane, the gaps are $s-$wave like with almost identical and nodeless within one pocket. Two distinct pairing symmetries are proposed: either the sign-reversed $s^{\pm}$-wave pairing mediated by inter-band repulsive interactions (driven by antiferromagnetic (AF) fluctuations) on the separated Fermi pockets \cite{korshunov2008s,avchubukov2008s,mazin2009s,Seo2008s,pjhirschfeld2011s,fwang2011s} or sign-preserved $s^{++}$ pairing mediated by inter-band attractive interactions (driven by orbital fluctuations) \cite{sonari2010s,hkontani2010s,sonari2012s,ltakeuchi2018s}. In the both scenarios, the pair-breaking energy is the sum of superconducting gaps on the hole and electron pockets connected by wavevector $\bf{Q}$: $\Delta_{tot}$ = $|\Delta_k|$ + $|\Delta_{k+Q}|$, which also defines the spin-flip continuum energy ($\hbar\omega_c$) of the spin-exciton under $s^{\pm}$-pairing as well \cite{tamaier2008s,tamaier2009s,tdas2011s,inosov2010s,mgkim2013s}. When $\Delta_k=\Delta_{k+Q}=\Delta$, $\hbar\omega_c=2\Delta$ is similar to that in cuprates \cite{meschrig2006s,ysidis2007s}.  If the sizes of hole and electron pockets are nearly the same, then the inter-band intra-orbital scattering will be enhanced for the Fermi surface nesting (weak-coupling picture) \cite{mazin2009s}. In this case, the neutron spin resonance induced by $s^{\pm}$-pairings is always commensurate at the wavevector $\bf{Q}$ linking the pair of hole-electron pockets (usually it is at $(\pi, \pi)$)). However, when the hole and electron pockets mismatch in sizes, thus Fermi surface nesting fails, the $s^{\pm}$-pairings mediated by short-range AF fluctuations (strong-coupling picture) \cite{Seo2008s,pjhirschfeld2011s,fwang2011s} can still induce a spin resonance \cite{tdas2011s}, but now it is incommensurate in reciprocal space with a incommensurability $d=\sqrt{|r_1^2-r_2^2|}$ determined by the the radiuses of hole and electron pocket $r_1$ and $r_2$ \cite{pdai2015s}.

In our previous study, high resolution Laser-Angle-Resolved-Photoemission-Spectroscopy (ARPES) measurements were performed on the single crystal samples of KCa$_2$Fe$_4$As$_4$F$_2$. The energy resolution was about 1 meV and the angular resolution was about 0.3$^{\circ}$ corresponding to 0.004 \AA$^{-1}$ momentum resolution at the photon energy of 6.994 eV. We have observed three dominate hole pockets ($\alpha,\beta,\gamma$) and one tiny electron pocket ($\delta$) with clearly bilayer splitting effects on $\beta$ and $\gamma$ bands \cite{dswu2020s}. The radius and superconducting gaps of these Fermi pockets are quite different, as shown in Fig. S9. The superconducting gaps can be described by a complex gap function, $\Delta_s=\mid \Delta_0(\cos k_x+\cos k_y)\pm\Delta_z[\cos(k_x/2)\cos(k_y/2)]\mid/2$, where $\Delta_0^{\alpha}=5.6$ meV, $\Delta_0^{\beta}=6.8$ meV, $\Delta_z^{\beta}=4$ meV, $\Delta_0^{\gamma}=3.3$ meV, $\Delta_z^{\gamma}=0.2$ meV and $\Delta_0^{\delta}=5.3$ meV.

For comparison, we have also performed Density Functional Theory (DFT) calculations with a plane wave basis set by using the experimental lattice parameters \cite{zcwang2016as}. Pseudo-potential and Plane-Wave Self-Consistent Field methods were used in our calculations based on Quantum Espresso code \cite{pgiannozzi2009s}. The plane wave cut off energy was set to 35 Ry after performing convergence test. Self-consistent-field (SCF) calculations were performed on 10$\times$10$\times$10 grid in $k$ space. The Fermi surfaces are estimated using CASTEP code, visualization tool "Fermi surfer" was used to show contribution of different orbitals, too \cite{sjclark2005s}. The cut off energy and $k$ mesh for SCF calculation were set as 500 eV and 26$\times$26$\times$37. The DFT results are presented in Fig. 1b, 1c, and Fig. S10.  The partial density of states (PDOS) at Fermi level is dominated by three orbitals: $d_{xz}$, $d_{yz}$ and $d_{xy}$ similar to other iron arsenides (Fig. S10b). Orbital projected band structure shows 6 hole-like bands around $\Gamma$ point and 4 electron-like bands around $M$ point (Fig. S10a). Consequently there are 6 hole pockets at Brillouin zone centre (the outer two bands are doubly degenerate) and 4 electron pockets around Brillouin zone corner (two of which are nearly degenerate), as marked in Fig. 1b, 1c and Fig. S10c (where the two degenerate bands are considered as one). Orbital character of the bands are shown with different color codes (Fig. S10d - S10h).

The superconducting gaps and radii of each Fermi pockets from ARPES measurement and DFT calculation results are summarized in Table S1. The total gaps $\Delta_{tot}$ and incommensurability $d$ of various scattering wavevector $Q \pm d$ are deduced from the data in Table S1 and listed in Table S2. For convenience, we have used the average size of electron pockets in DFT results ($r_2=0.24 \pi/a$). In both cases, the values of $d$ are similar, and $\Delta_{tot}$ decreases when $d$ increases, as shown in Fig. 4a.
\\

\begin{table}[h]
Table S1. Superconducitng gaps and Fermi pocket sizes in KCa$_2$Fe$_4$As$_4$F$_2$.\\
\begin{center}
\begin{tabular}{|c|c|c|c|c|c|c|c|c|}
 \hline \hline
pocket & $\alpha$ & $\beta_1$ & $\beta_2$ & $\gamma_1$ & $\gamma_2$ & $\delta_1$ & $\delta_2$ & $\delta_3$ \\
  \hline
ARPES Gap (meV)    & 5.6       & 4.1     & 7.9      & 2.1   &  1.9  & 5.3 & - & - \\
\hline
DFT radii ($\pi$/a)     & 0.14        & 0.38      & - & 0.47  & 0.49   & 0.18 & 0.24 & 0.29 \\
\hline
ARPES radii ($\pi$/a)    & 0.10       & 0.22      & 0.23  & 0.43   & 0.44   & 0.04 & - & -  \\
\hline \hline
\end{tabular}
\end{center}\label{tab:table 1}
\end{table}

\begin{table}[h]
Table S2. $|\Delta_{tot}|$ and $d$ in KCa$_2$Fe$_4$As$_4$F$_2$.\\
\begin{center}
\begin{tabular}{|c|c|c|c|c|c|}
 \hline \hline
scattering vectors & $\alpha \rightarrow \delta$ & $\beta_1 \rightarrow \delta$  & $\beta_2 \rightarrow \delta$ & $\gamma_1 \rightarrow \delta$ & $\gamma_2 \rightarrow \delta$ \\
  \hline
ARPES $\Delta_{tot}$ (meV)    & 10.9      & 9.4     & 13.2      & 7.4   &  7.2 \\
\hline
ARPES $d$ (r.l.u)     & 0.046        & 0.108      & 0.113 & 0.214  & 0.219 \\
\hline
DFT  $d$ (r.l.u)   & 0.097       & 0.144      & -  & 0.201   & 0.216  \\
\hline \hline
\end{tabular}
\end{center}\label{tab:table 2}
\end{table}

\newpage
\begin{figure*}[t]
\renewcommand\thefigure{S1}
\includegraphics[width=0.8\textwidth]{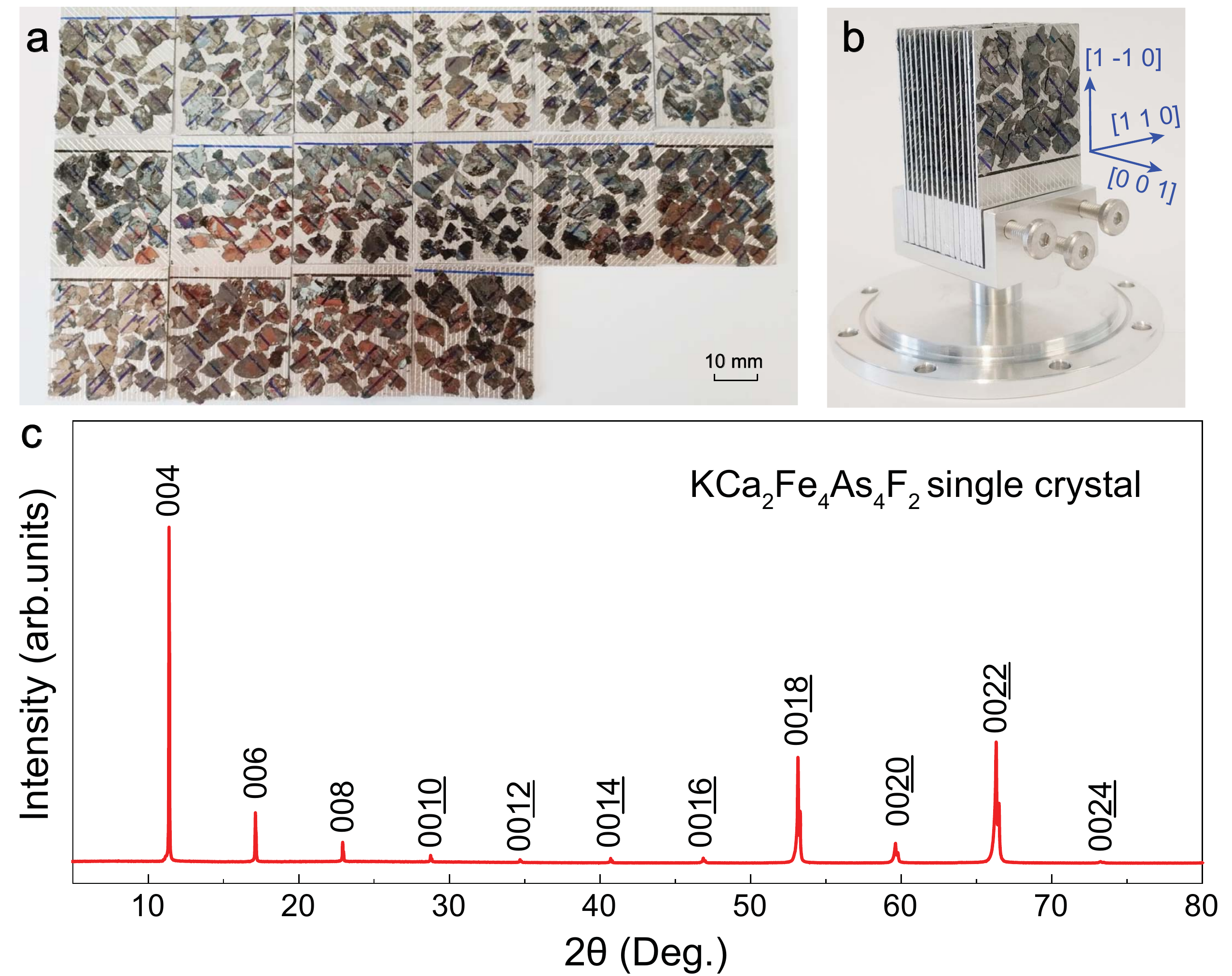}
\caption{ {\bf Photos and X-ray diffraction patterns of KCa$_2$Fe$_4$As$_4$F$_2$ single crystals.}
{\bf a}, Co-aligned crystals on aluminum plates for neutron scattering experiments.
{\bf b}, Sample mount in the [$H, H, L$] scattering plane.
{\bf c}, X-ray diffraction patterns with incident beam along $c-$axis.
}
\end{figure*}

\begin{figure*}[t]
\renewcommand\thefigure{S2}
\includegraphics[width=0.8\textwidth]{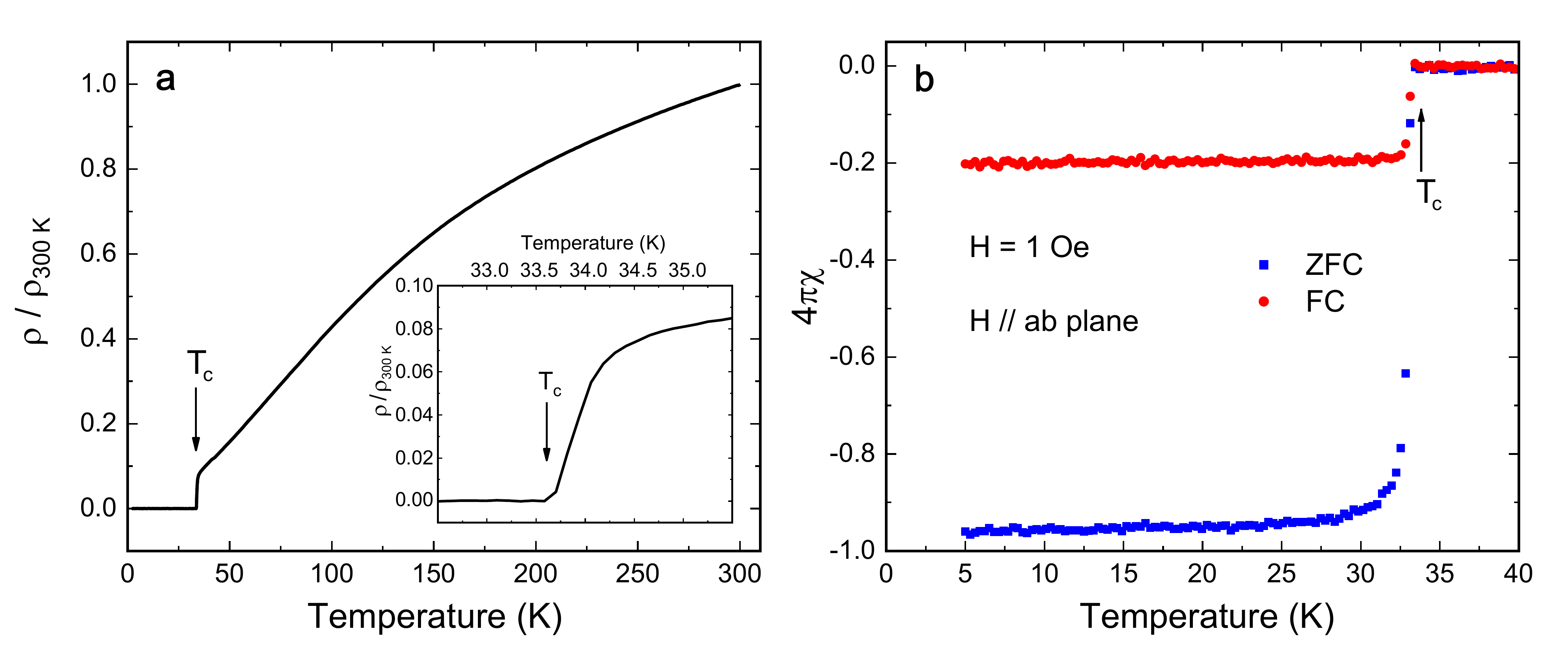}
\caption{
{\bf Superconducting properties of KCa$_2$Fe$_4$As$_4$F$_2$ crystals}.
{\bf a}, Temperature dependence of the resistivity within the $ab-$plane, the data is normalized by the resistivity at 300 K. Insert shows the superconducting transition at $T_c=33.5$ K.
{\bf b}, Magnetic susceptibility measurements by zero-field-cooling (ZFC) and field-cooling (FC) methods with $H // ab$.
}
\end{figure*}

\begin{figure*}[t]
\renewcommand\thefigure{S3}
\includegraphics[width=0.8\textwidth]{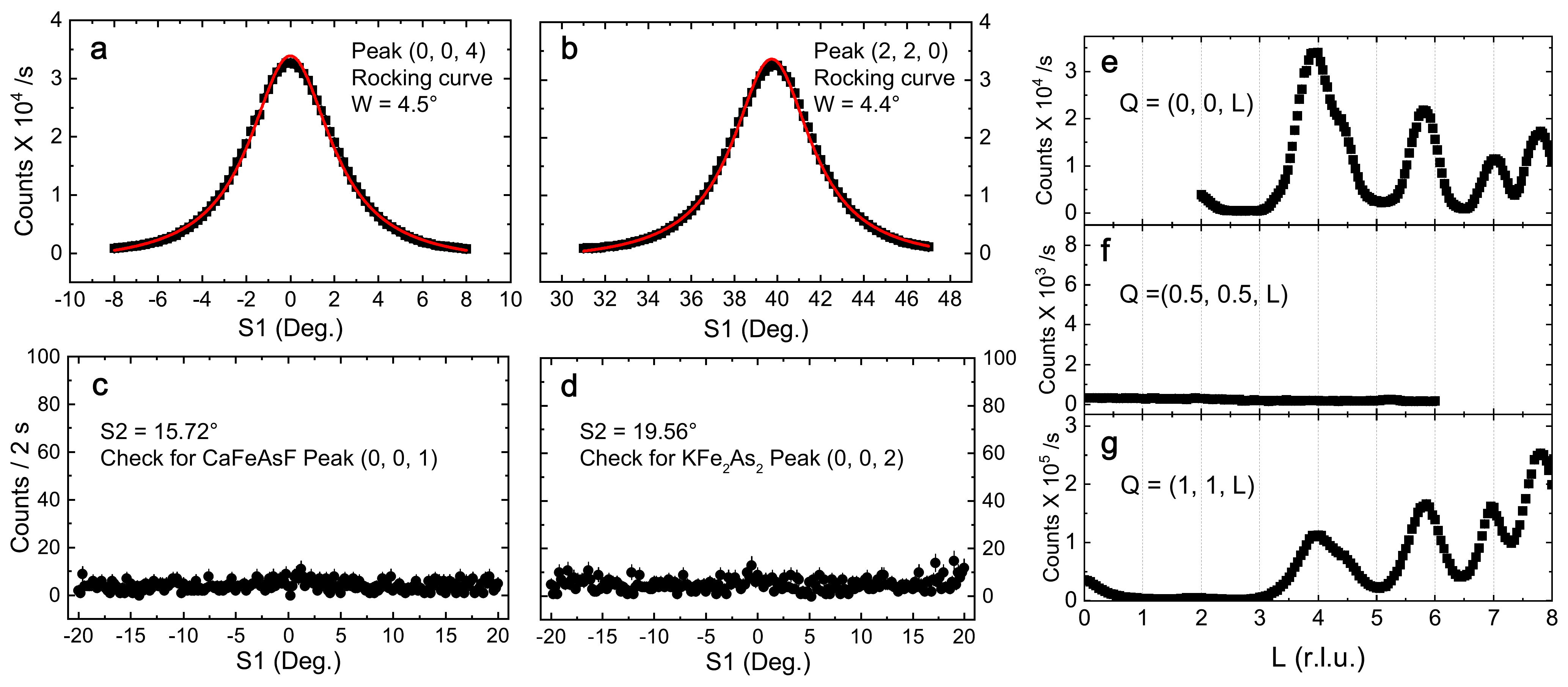}
\caption{
{\bf Elastic neutron scattering measurements for sample alignments and impurity checks.}
 {\bf a, b}, Rocking curves of Bragg peaks at $(0, 0, 4)$ and $(2, 2, 0)$.
 {\bf c, d}, Rocking curves for checking on possible impurity phases of KFe$_2$As$_2$ and CaFeAsF.
 {\bf e, f, g}, Long $L$ scans along $Q$ = $[0, 0, L]$, $Q$ = $[0.5, 0.5, L]$ and $Q$ = $[1, 1, L]$.
}
\end{figure*}

\begin{figure*}[t]
\renewcommand\thefigure{S4}
\includegraphics[width=0.9\textwidth]{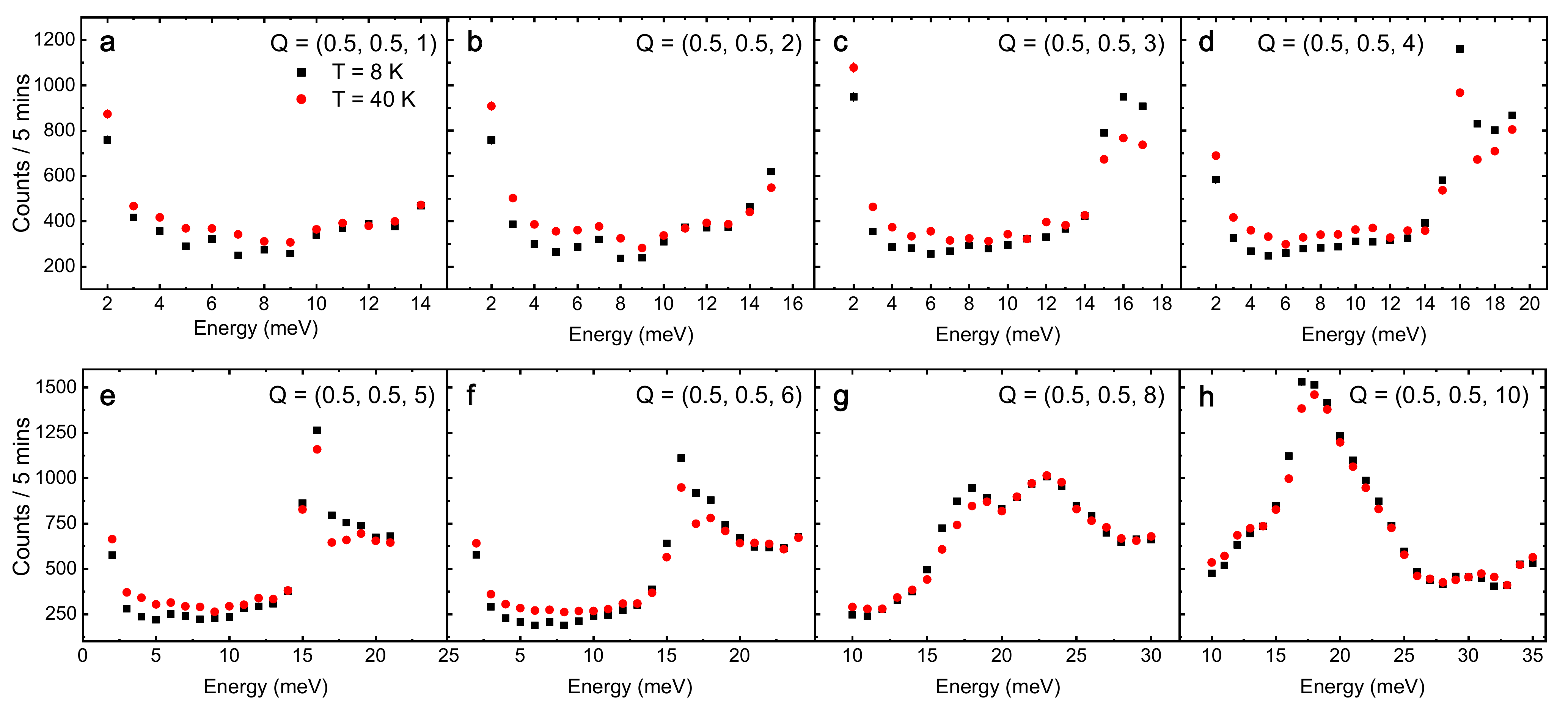}
\caption{
{\bf Energy scans at $Q = (0.5, 0.5, L)$ ($L$ = 1, 2, 3, 4, 5, 6, 8) below (black) and above (red) $T_c$.}
}
\end{figure*}

\begin{figure*}[t]
\renewcommand\thefigure{S5}
\includegraphics[width=0.8\textwidth]{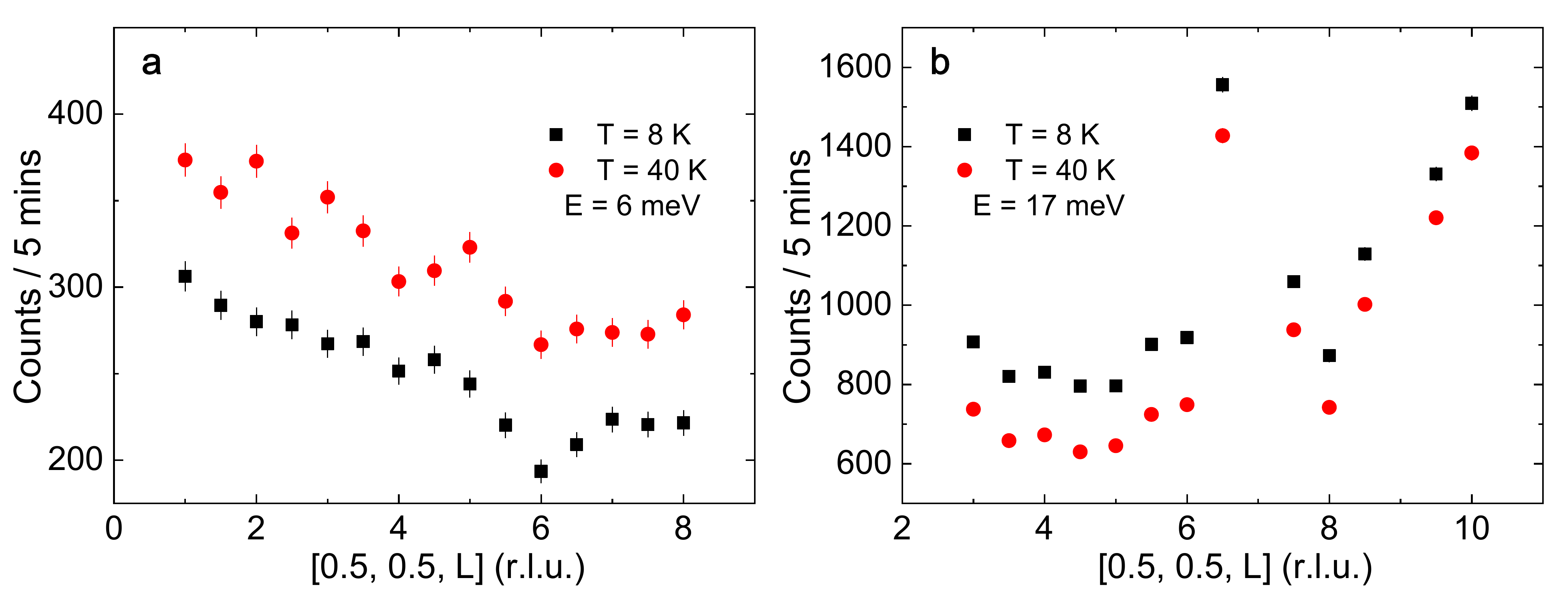}
\caption{
{\bf Constant-energy scans ($Q-$scans) along $Q=[0.5, 0.5, L]$ direction at $E$ = 6 meV and $E$ = 17 meV.}
}
\end{figure*}

\newpage
\begin{figure*}[t]
\renewcommand\thefigure{S6}
\includegraphics[width=0.95\textwidth]{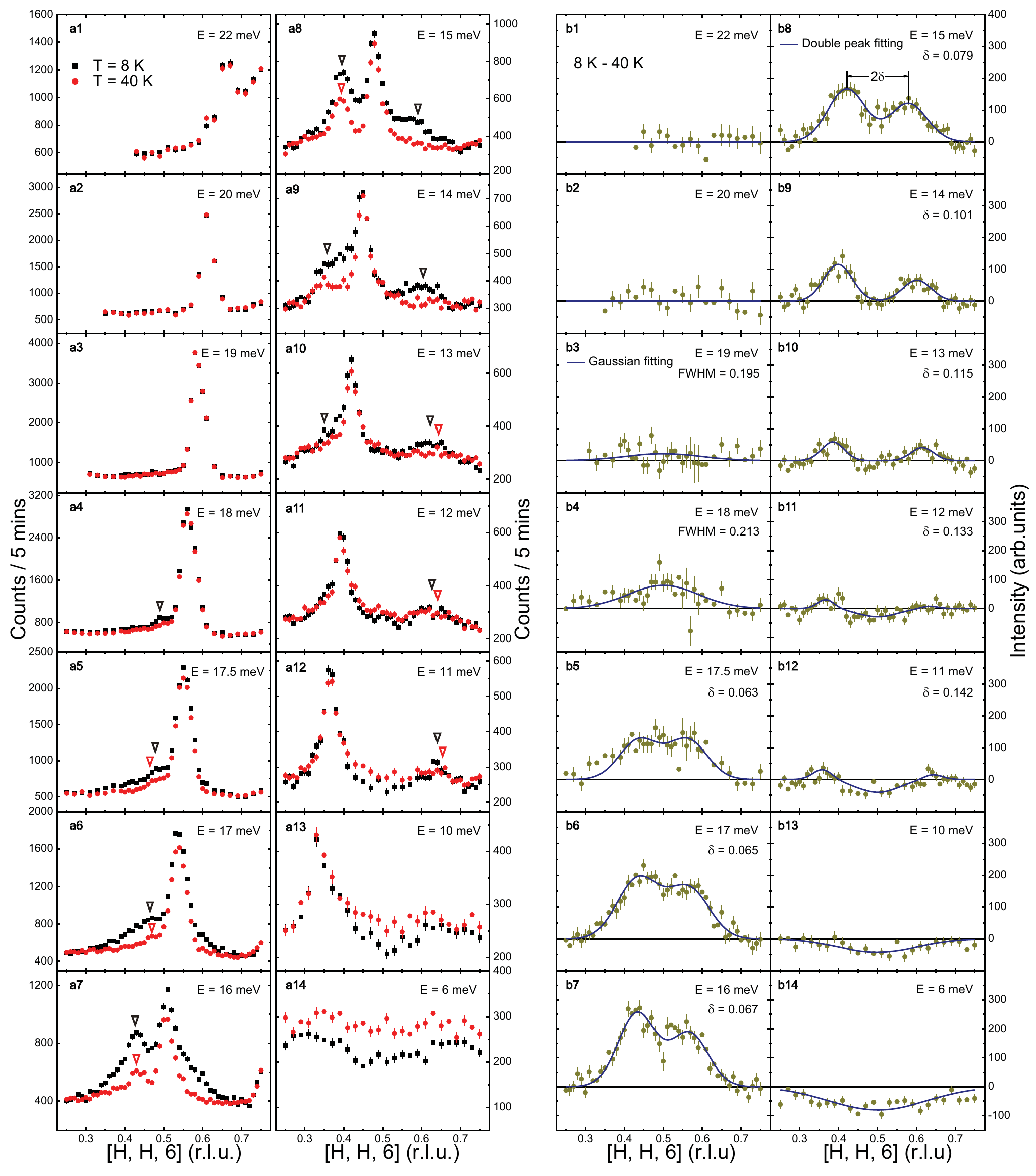}
\caption{
{\bf Constant-energy scans along $Q=[H, H, 6]$ and their differences for $E$ = 6, 10, 11, 12, 13, 14, 15, 16, 17, 17.5, 18, 20 and 22 meV}. The triangles mark the incommensurate peak of the magnetic scattering, and the solid line are gaussian fits by one or double functions. The incommensurability $\delta$ is defined by the distance between two peaks as shown in b5 - b12.
}
\end{figure*}

\begin{figure*}[t]
\renewcommand\thefigure{S7}
\includegraphics[width=0.8\textwidth]{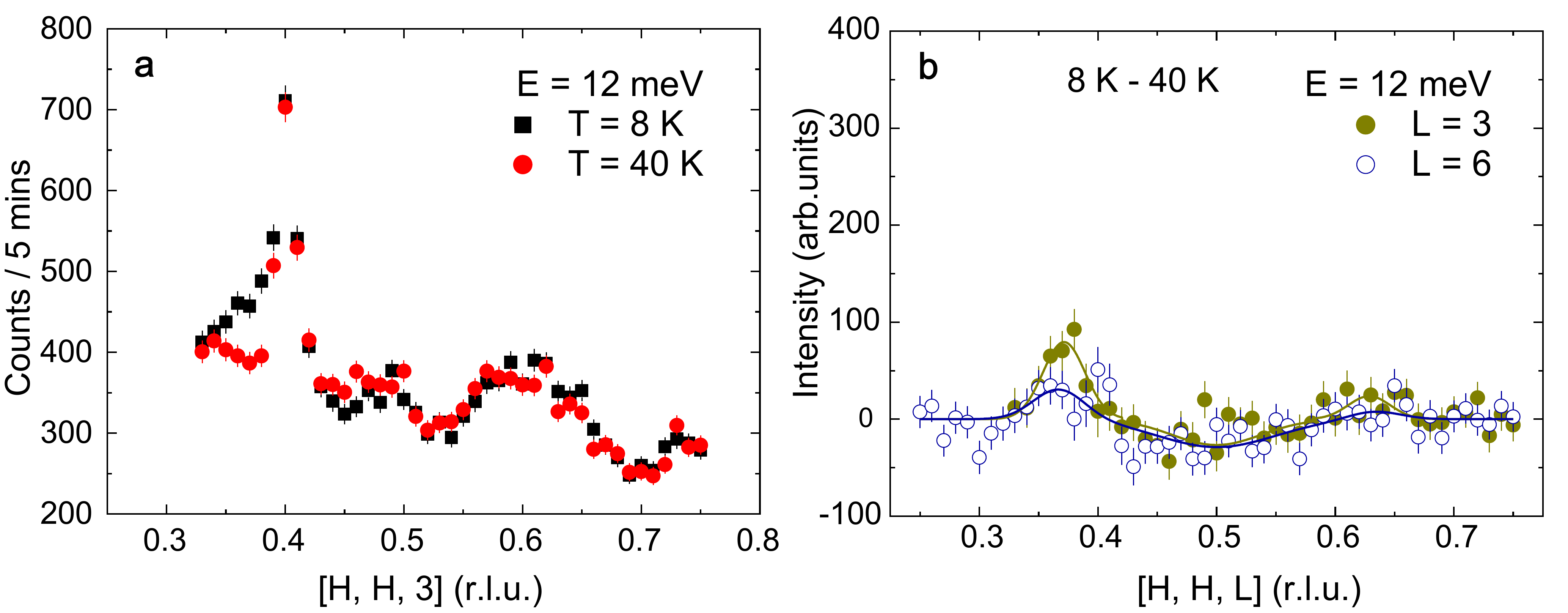}
\caption{
{\bf Constant-energy scans along $Q=[H, H, 3]$ and $[H, H, 6]$ for $E$ = 12 meV.}
{\bf a}, Raw data of constant-energy scans below and above $T_c$ along $Q=[H, H, 3]$.
{\bf b}, Comparison of incommensurate resonant peaks distribution for $[H, H, 3]$ and $[H, H, 6]$ scans.
}
\end{figure*}

\begin{figure*}[t]
\renewcommand\thefigure{S8}
\includegraphics[width=0.8\textwidth]{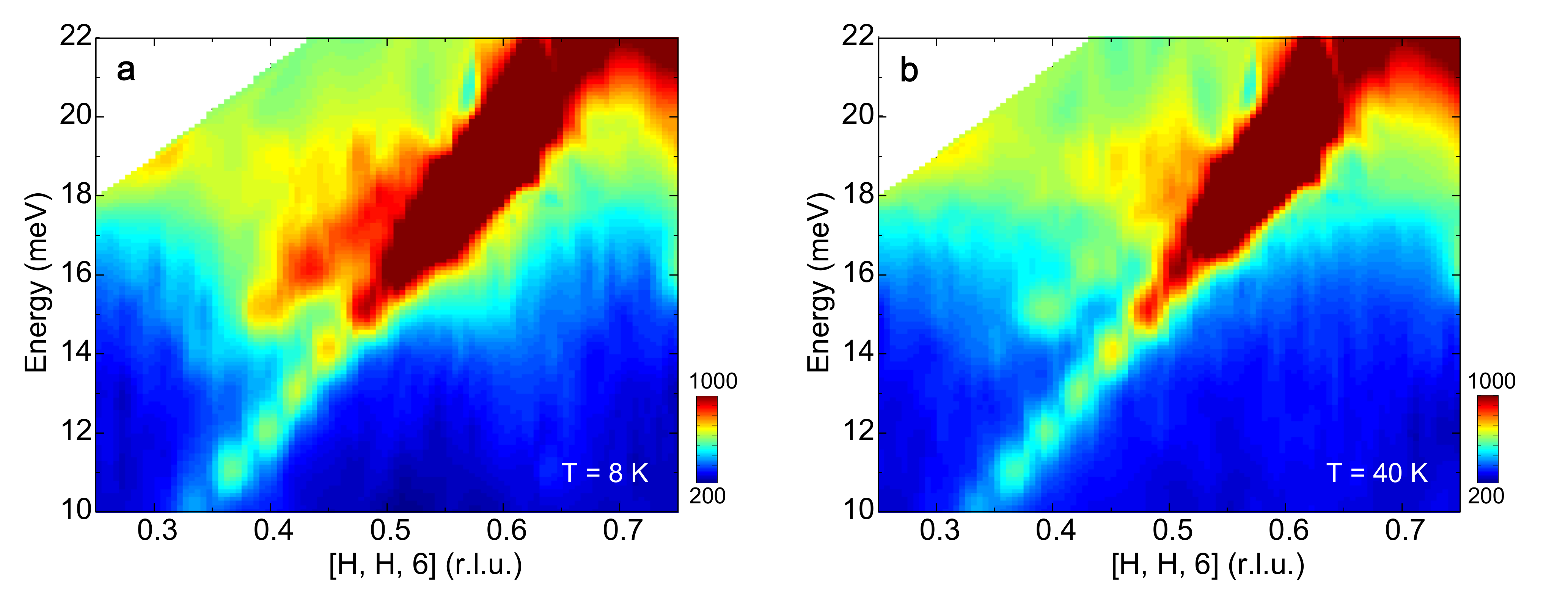}
\caption{
{\bf Color mapping of the raw data of all $Q-$scans along $[H, H, 6]$ with $E$ = 10 $\sim$ 22 meV below and above $T_c$ corresponding to Fig. S6 a1 - Fig. S6 a14}
}
\end{figure*}

\begin{figure*}[t]
\renewcommand\thefigure{S9}
\includegraphics[width=0.6\textwidth]{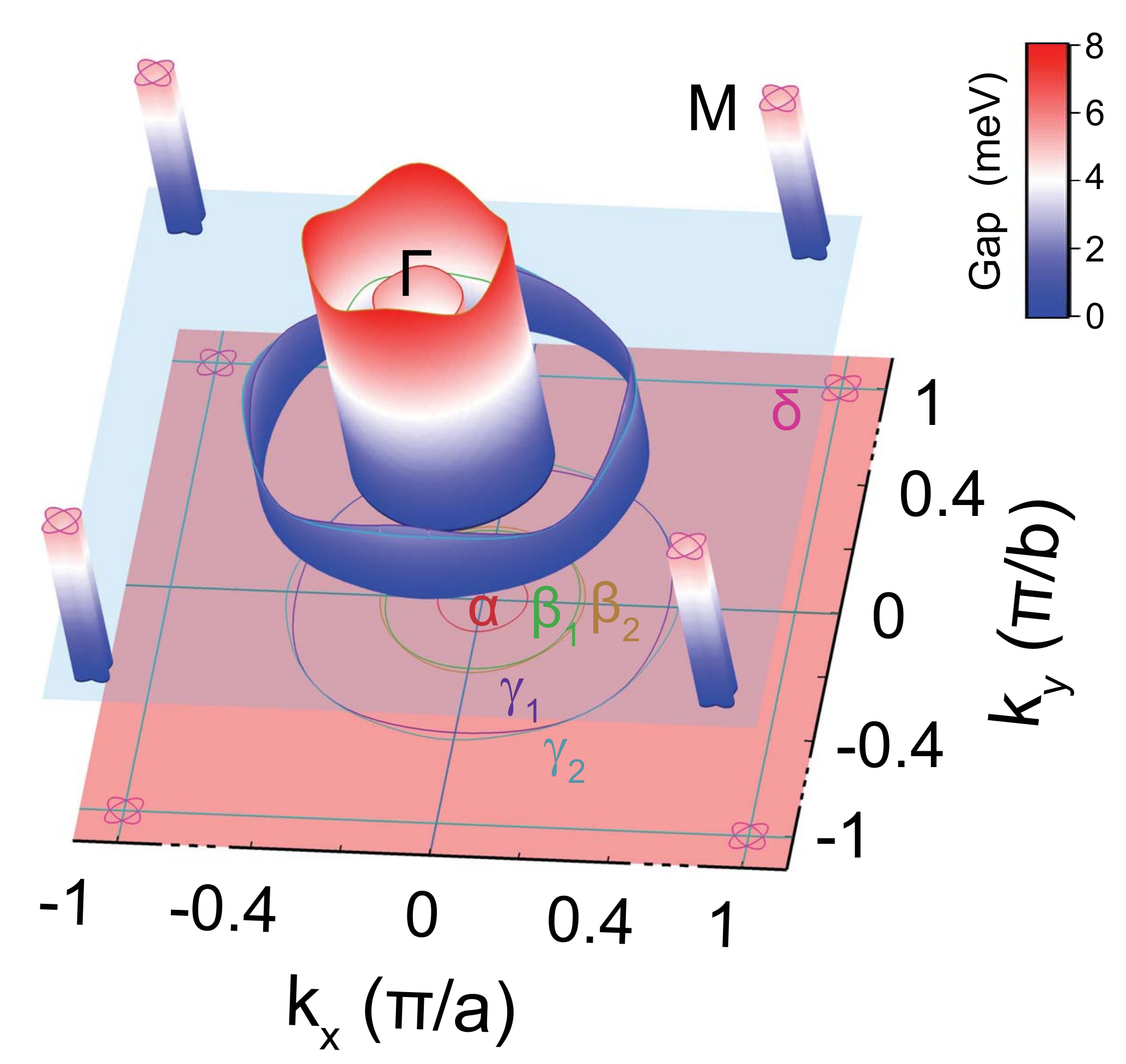}
\caption{
{\bf ARPES results for the Fermi surfaces and superconducting gaps.}
}
\end{figure*}

\begin{figure*}[t]
\renewcommand\thefigure{S10}
\includegraphics[width=0.9\textwidth]{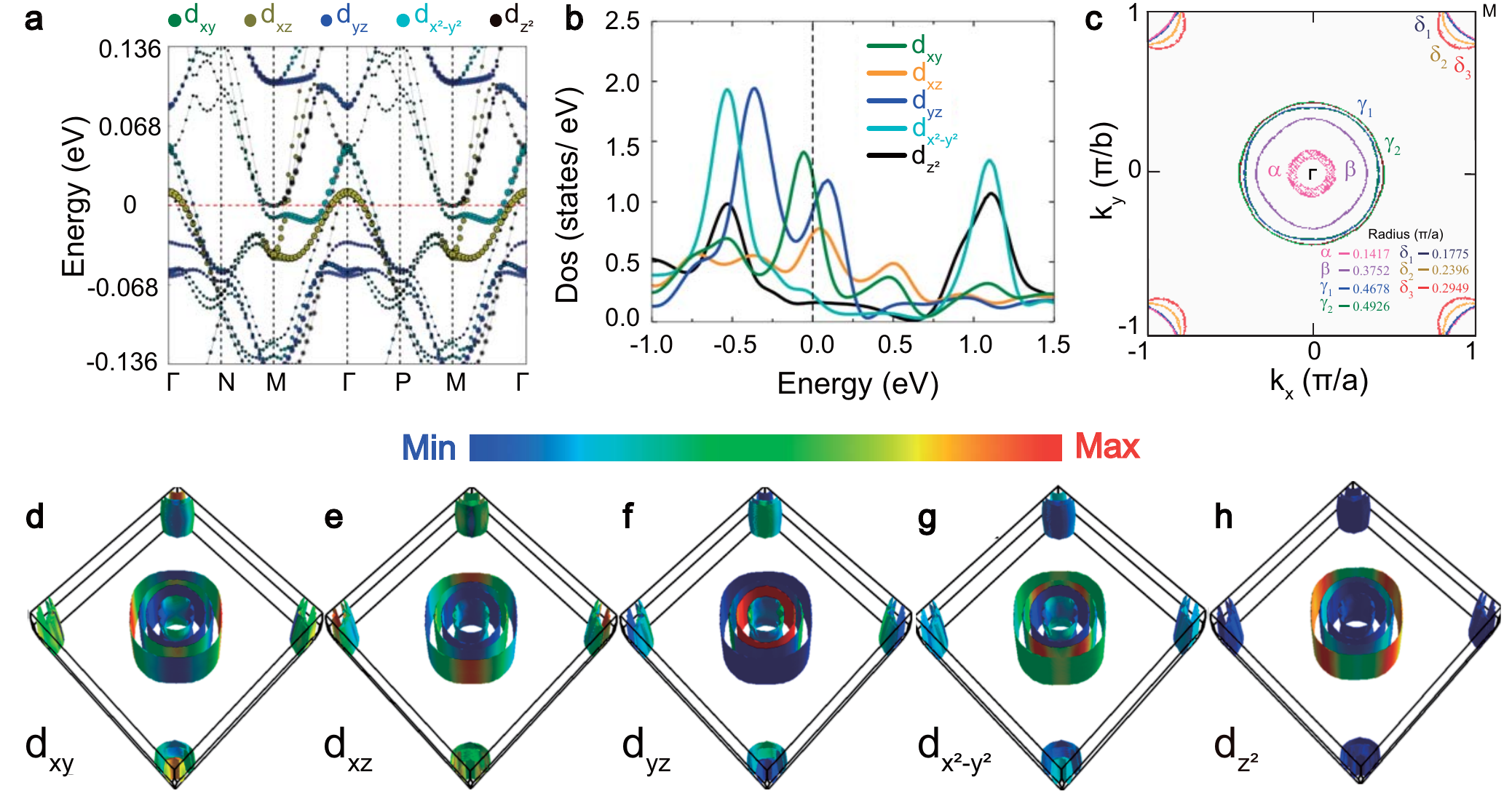}
\caption{
{\bf DFT calculation results for the band structure and Fermi surfaces.}
{\bf a}, Electronic band structure and its orbital distributions.
{\bf b}, Partial-density-of-state (PDOS) for each orbital of Fe$^{2+}$.
{\bf c}, Fermi surfaces and their sizes.
{\bf d - h}, Orbital characters of the Fermi surfaces shown with different color codes.
}
\end{figure*}

\end{document}